\documentclass{PoS}

\title{More discoveries of compact radio cores in Seyfert galaxies with the EVN}

\ShortTitle{More discoveries of compact radio cores in Seyfert galaxies with the EVN}

\author{\speaker{Marcello Giroletti}\\
        INAF Istituto di Radioastronomia\\
	via Gobetti 101, 40129 Bologna (Italy) \\
        E-mail: \email{giroletti@ira.inaf.it}}

\author{Francesca Panessa\\
        INAF IASF, via Fosso del Cavaliere 100, 00133 Rome, Italy\\
        E-mail: \email{francesca.panessa@iasf-roma.inaf.it }}

\abstract{Following the high detection rate achieved by EVN observations of the central regions of local Seyfert galaxies (Giroletti \& Panessa 2009, ApJL 706, 260, \cite{Giroletti2009}), we have targeted a few additional sources from a complete sample. We have detected three more sources (NGC 3982, NGC 3227, and NGC 4138) at both 1.6 and 5 GHz and present preliminary results. Moreover, the declination of the sources was suitable to include Arecibo in the EVN observations, which provides important clues on the compactness of the emission region.}

\FullConference{10th European VLBI Network Symposium and EVN Users Meeting: VLBI and the new generation of radio arrays\\
		September 20-24, 2010\\
		Manchester Uk}

\begin{document}

\section{Background: recent VLBI observations of local Seyfert galaxies}

\begin{figure}
\center \includegraphics[width=0.8\textwidth]{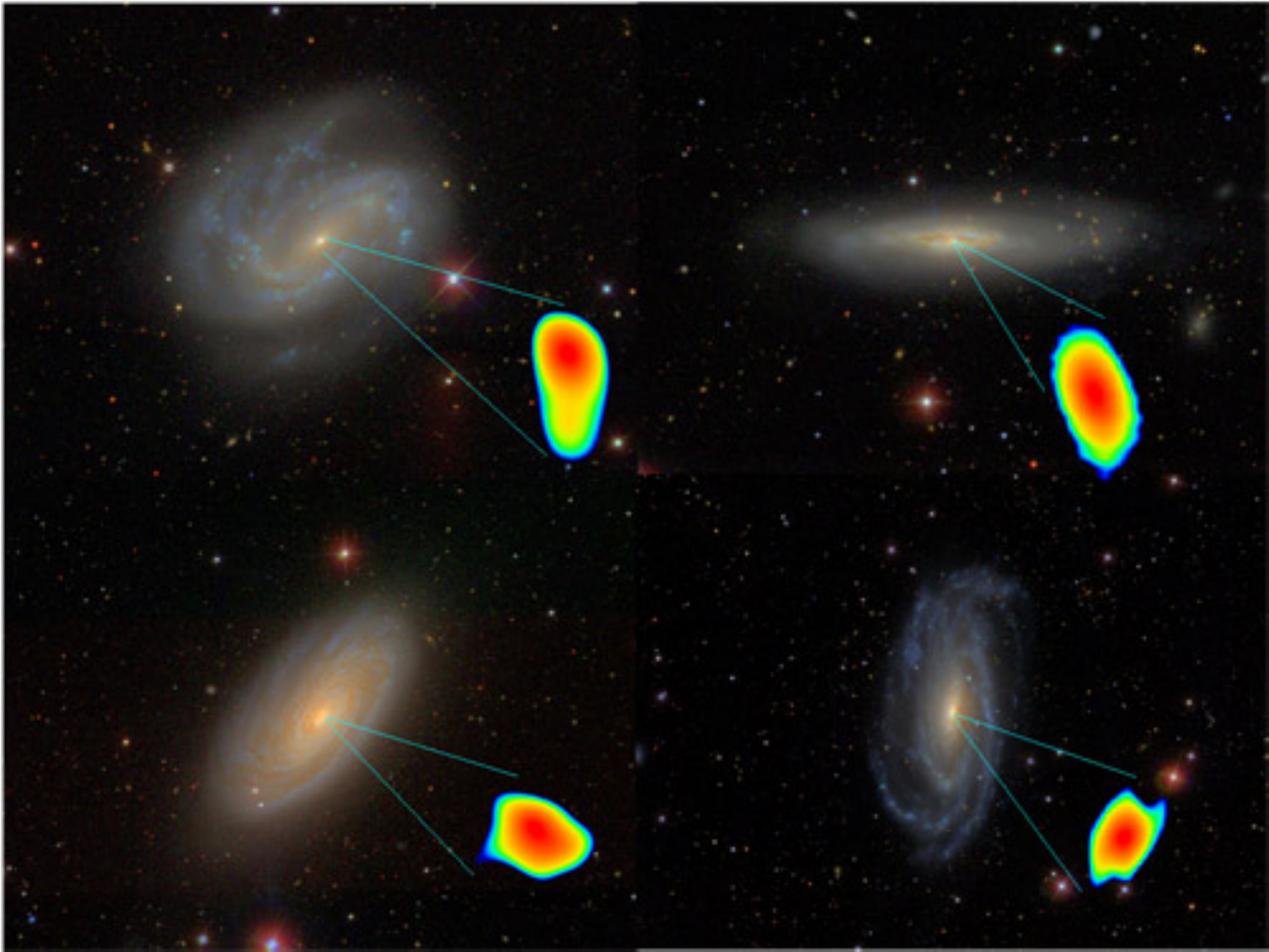}
  \caption{Combined optical and EVN images of Seyfert galaxies. Clockwise from top left: NGC\,4051, NGC\,4388, NGC\,5033, and NGC\,4501. Optical images from Sloan Digital Sky Survey, radio ones from EVN observations at 1.6 GHz (\cite{Giroletti2009}, typical beam $10\times5$ mas). \label{f.2009}}
\end{figure}

It is a well established fact that radio quiet Active Galactic Nuclei (AGNs) are not radio silent:\ indeed, VLA observations reveal emission at milliJansky level in most AGNs, even in low-luminosity AGNs \cite{Nagar2002}. It is then natural to wonder about the nature of this emission. In particular, it is relevant to understand whether it is compact or diffuse. We are carrying out an extensive analysis of the high resolution radio properties of a complete distance limited ($d<22\,$Mpc) sample of local Seyfert galaxies \cite{Cappi2006}, both from the literature and new observations. We find evidence in the literature for the presence of compact emission from published VLBI observations of the brightest members of the sample \cite{Gallimore2004,Ulvestad2005,Wrobel2006}; however, it was necessary to request new dedicated observations for the majority of the weak sources. As a start, we observed 5 nuclei of Seyfert galaxies with the EVN from this sample, and successfully detected compact cores in  4 of them (Fig.~\ref{f.2009}, \cite{Giroletti2009}).

The observational and physical properties determined for such sources were however quite different from target to target, with the presence of both compact and resolved structures (in one case even with multi-components) and of flat and steep spectra.

\section{New observations}

The sources observed up to \cite{Giroletti2009} still made up a rather biased subset of the original sample, as observations of the faintest sources were still missing. In total, 13 of them have not been observed with VLBI yet; 8 of them have a VLA detection in at least one frequency \cite{Ho2001}. Therefore, we performed new 1.6 and 5 GHz observations for these 8 targets. The observations were carried out with the EVN in June 2009, with 1 Gbps recording rate. The partecipating telescopes were Jodrell Bank, Westerbork, Effelsberg, Onsala, Medicina, Torun, Noto, plus the 300m Arecibo radio telescope for the sources of suitable declination and the new 40m Yebes telescope at 5 GHz.

\section{Preliminary results}

\begin{figure}
\center \includegraphics[width=0.8\textwidth]{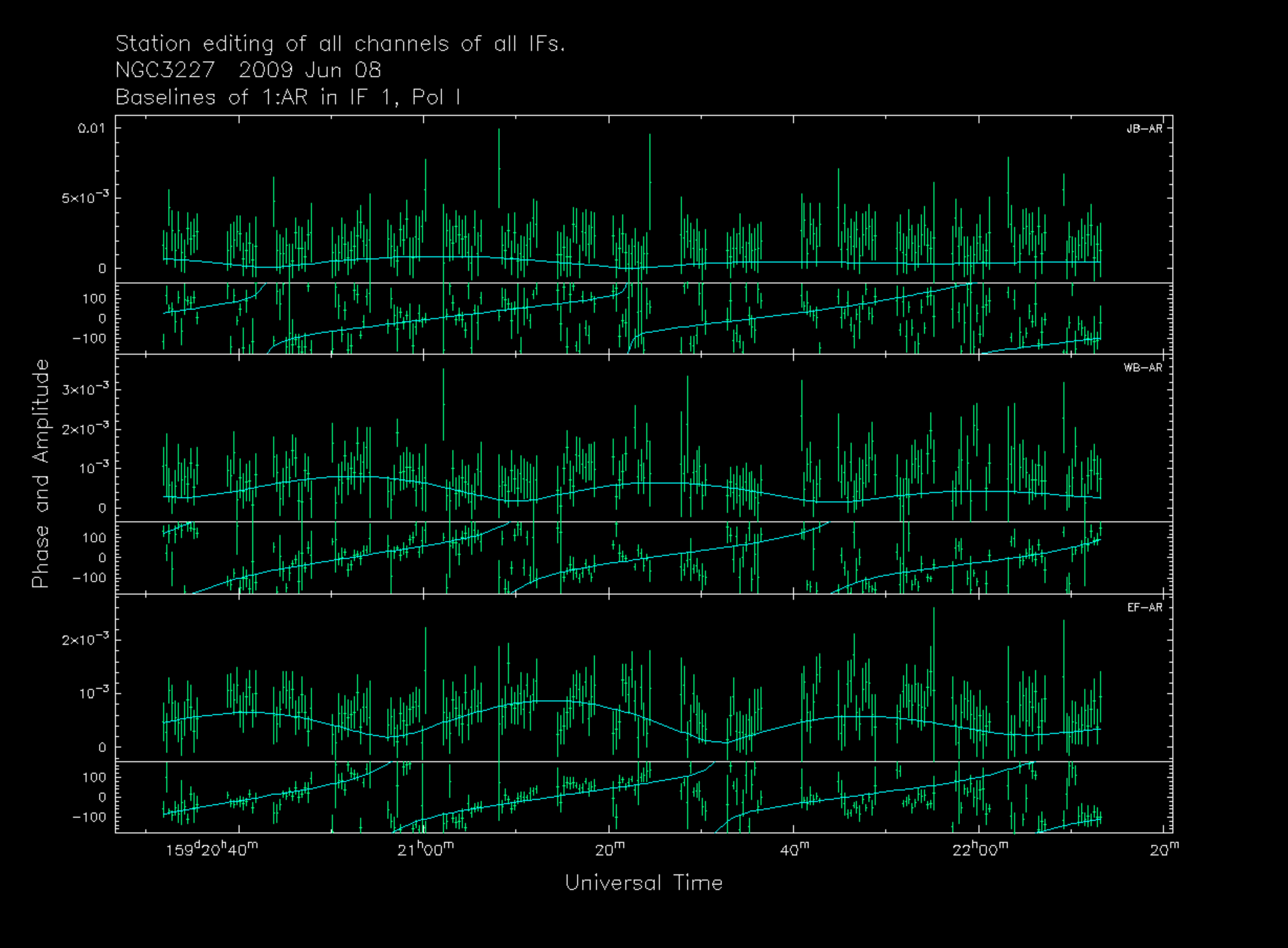}
  \caption{Visibility amplitude and phase vs.\ time for baselines from Arecibo to Jodrell Bank (top panels), Effelsberg (middle), and Westerbork (bottom). The target source is NGC\,3227. \label{f.vplot}}
\end{figure}

At a first look, based on pipeline calibrated data, three sources are clearly detected in at least one of the two bands, and namely NGC\,3982, NGC\,3227, NGC\,4138. The cleaned flux density is at the mJy level, with rms $\sim 30 \, \mu$Jy. The addition of Arecibo is particularly relevant, as it provides sensitive long baselines. The observations of the five targets in Giroletti \& Panessa (2009) were very sensitive but somewhat resolution limited; indeed, phase coherence was best detected on the sensitive but short baselines between Effelsberg, Jodrell Bank, and Westerbork. We now show in Fig.~\ref{f.vplot} the visibility amplitudes and phases on baselines to Arecibo for one of the newly observed sources (NGC\,3227). Phase coherence is clearly present in these transatlantic baselines, which proves that the target is truely compact. For instance, in the case of NGC\,3227 at 5 GHz, a preliminary visibility model fit indicates that the source is at least as compact as $0.19 \times 0.09\,$pc.

Albeit preliminary, these results confirm that compact radio emission is present in most galactic nuclei, even if their broad band luminosity is very low (as small as $10^{19}\,$W$\,$Hz$^{-1}$ in the radio). It will be now possible to revisit the connection to the X-ray properties (originally explored e.g.\ by Panessa et al.\ 2007, \cite{Panessa2007}) making use of high resolution data. This will permit us to obtain a better understanding of the exact location of the radio emission and of the interplay between the physical mechanisms responsible for the radio and X-ray emission.

\section*{Acknowledgments}
We acknowledge financial contribution from ASI-INAF I/088/06/0. The European VLBI Network is a joint facility of European, Chinese, South African and other radio astronomy institutes funded by their national research councils.

\end{document}